\documentclass[runningheads]{llncs}
\usepackage[T1]{fontenc}
\usepackage{graphicx}
\usepackage{xcolor}
\usepackage{comment}
\usepackage{booktabs}
\usepackage{tcolorbox}
\definecolor{tcolorbox}{gray}{0.90}
\usepackage{eso-pic}
\usepackage{url}
\usepackage[backend=biber,style=numeric,sortcites]{biblatex}
\providecommand{\orcidID}[1]{}

\begin{document}


\AddToShipoutPictureFG*{
  \AtPageUpperLeft{
    \put(0.5\paperwidth, -8mm){
      \makebox(0,0)[c]{
        \fbox{\parbox{8.5cm}{\centering\normalsize\textbf{A version of this paper appears in PAM 2026\footnotemark[1]}}}
      }
    }
  }
}
\footnotetext[1]{\url{https://doi.org/10.1007/978-3-032-18268-5_15}}

\title{State of Passkey Authentication in the Wild: A Census of the Top 100K sites}

\author{Prince Bhardwaj\, \orcidID{0009-0004-6434-335X} \and
Nishanth Sastry\, \orcidID{0000-0002-4053-0386}}

\institute{University of Surrey, United Kingdom
\email{\{p.pawankumarsharma,n.sastry\}@surrey.ac.uk}}
\maketitle

\begin{abstract}
Passkeys -- discoverable WebAuthn credentials synchronized across devices are widely promoted as the future of passwordless authentication. Built on the FIDO2 standard, they eliminate shared secrets and resist phishing while offering usability through platform credential managers. Since their introduction in 2022, major vendors have integrated passkeys into operating systems and browsers, and prominent websites have announced support. Yet the true extent of adoption across the broader web remains unknown.

Measuring this is challenging because websites implement passkeys in heterogeneous ways. Some expose explicit ``Sign in with passkey'' buttons, others hide options under multi-step flows or rely on conditional mediation, and many adopt external mechanisms such as JavaScript libraries or OAuth-based identity providers. There is no standardized discovery endpoint, and dynamic, JavaScript-heavy pages complicate automated detection.

This paper makes two contributions. First, we present \emph{Fidentikit}, a browser-based crawler implementing 43 heuristics across five categories -- UI elements, DOM structures, WebAuthn API calls, network patterns, and library detection developed through iterative refinement over manual examination of 1{,}500 sites. Second, we apply Fidentikit to the top 100{,}000 Tranco-ranked domains, producing the first large-scale census of passkey adoption. Our results show adoption strongly correlates with site popularity and often depends on external identity providers rather than native implementations.
\end{abstract}
$\textbf{Keywords:}$ passkeys, WebAuthn, FIDO2, passwordless, large‑scale measurement, authentication.

\section{Introduction}

Passwords remain the dominant form of web authentication, yet they are inherently insecure. Users frequently reuse passwords across services, enabling credential stuffing attacks. Even unique passwords are vulnerable when websites fail to store them securely, exposing plaintext or weakly hashed credentials in breaches. Worse, a stolen password can be replayed indefinitely unless detected and revoked. These weaknesses have driven the search for a ``holy grail'' of passwordless authentication~\cite{stajano2011pico}: mechanisms that eliminate shared secrets, resist phishing, and reduce reliance on human memory.

Passkeys represent the most promising realization of this vision. A passkey is a discoverable WebAuthn credential synchronized across a user’s devices via encrypted cloud storage, enabling seamless multi-device login without passwords. Built on the FIDO2 standard, passkeys inherit the phishing resistance of public-key cryptography: private keys never leave the authenticator, and signatures are origin-bound, preventing replay on lookalike domains. Unlike device-bound FIDO credentials tied to a single hardware token, passkeys leverage platform credential managers such as Apple iCloud Keychain, Google Password Manager, and Microsoft Authenticator to provide ubiquity and convenience. Introduced in 2022 by the FIDO Alliance~\cite{FIDOAllianceFIDO2}, passkeys have since been championed by major vendors and integrated into mainstream operating systems and browsers. Industry announcements tout rapid adoption~\cite{NCSC2025_passkey}, yet these claims remain anecdotal and skewed toward high-profile brands. The true extent of passkey deployment across the broader web ecosystem is unknown.

Passkeys have matured beyond early prototypes: platform support is stable, major websites have rolled out passkey login options, and developer tooling has improved. At the same time, the ecosystem faces fragmentation. Unlike OAuth or OpenID Connect, which expose standardized discovery endpoints, WebAuthn lacks machine-readable metadata for passkey support. A site may implement passkeys without advertising this capability in any predictable URL or API. Consequently, automated detection is non-trivial. Websites differ in how they surface passkey options: some display explicit ``Sign in with a passkey'' buttons, others embed them under ``More sign-in options,'' and still others rely on conditional mediation, where the browser reveals passkey choices only after user interaction. Non-English sites introduce further variability with localized terminology. Beyond UI heterogeneity, technical signals vary: some sites invoke WebAuthn APIs immediately on page load, others defer calls until after multi-step flows. These challenges render na\"ive crawling ineffective.

Our goal in this paper is twofold. First, we present a robust, reproducible methodology for detecting passkey support at scale. We developed \textbf{Fidentikit}, an open-source browser-based crawler that extends SSO-Monitor~\cite{sso-monitor}, equipping it with heuristics that capture the diverse ways websites implement passkeys. Our system combines five classes of heuristics: (1) UI text and ARIA labels, (2) DOM attributes and form structures, (3) JavaScript WebAuthn API invocations, (4) network request patterns, and (5) known WebAuthn library filenames. These heuristics were derived through an iterative, semi-automated process. Starting with manual inspection of the top 100 Tranco-ranked sites, we encoded observed patterns into detection rules, validated them on progressively larger sets (ranks 101--500, then 501--1{,}000, and so on), and refined the rules whenever false negatives emerged. After examining over 1{,}500 sites, the marginal discovery rate fell below 2\%, yielding a primary set of 43 heuristics which we group into five classes described above. This system forms our first contribution: an open-source crawler and passkey detection framework hosted at~\url{https://netsys.surrey.ac.uk/softwares/fidentikit/} and Admin Dashboard shown in Figure~\ref{fig:dashboard} of Appendix~\ref{appendix:regex}, enabling researchers and practitioners to verify passkey support and identify the heuristic that triggered detection.

Second, we apply this crawler to conduct the first large-scale census of passkey adoption in the wild. Using the Tranco list, we measured passkey support across the top 100{,}000 websites. Our analysis reveals several key findings. First, adoption is strongly correlated with site popularity: top-ranked domains exhibit substantially higher passkey support than lower ranked sites. Second, many sites do not implement passkeys natively but rely on external mechanisms. Two dominant patterns emerge: (a) integration of JavaScript libraries such as \texttt{fido-lib}, \texttt{webauthn-framework}, and \texttt{passwordless.id}, and (b) offering OAuth-based login via identity providers like Google, which themselves support passkeys. In the latter case, a user can authenticate to the site using OAuth and complete the login with a passkey at the identity provider, effectively extending passkey benefits without direct implementation by the relying party. We also find that passkey adoption varies by geographic region, with sites hosted in the USA, Europe and Russia having the highest passkey adoption rates, followed by countries like Australia and India.

In summary, this paper makes the following contributions:
\begin{enumerate}
    \item \textbf{Fidentikit}, a scalable tool and crawler for heuristics-based passkey detection that adapts SSO-Monitor~\cite{sso-monitor} for passkey logins. Fidentikit adds 43 heuristics across five categories to enable reliable identification of passkey support on modern, JavaScript-heavy websites.
    \item A comprehensive census of passkey adoption across the top 100K Tranco-ranked domains, providing empirical insight into deployment trends, implementation strategies, and reliance on external identity providers.
\end{enumerate}

Our results offer a grounded perspective on the state of passwordless authentication on the web, complementing prior usability and protocol-level studies with quantitative evidence from the field. By illuminating where and how passkeys are deployed, we aim to inform both researchers and practitioners seeking to accelerate the transition toward a passwordless future.

\noindent\textbf{Scope and Limitations.} We emphasise that this work focuses on measuring passkey \textit{deployment}, whether websites offer passkey authentication as an option rather than measuring actual \textit{user adoption} of passkeys. Understanding user-side adoption would require access to authentication telemetry or user surveys, which is beyond the scope of this measurement study. Nevertheless, quantifying service-side availability is a prerequisite for understanding adoption barriers: users cannot adopt passkeys if websites do not offer them. Our findings reveal that whilst major service providers have adopted passkeys, they are predominantly offered as an \textit{additional} authentication method alongside passwords and traditional MFA, rather than as a replacement. This observation suggests that the ``passwordless future'' remains aspirational, with passkeys currently serving as a supplementary security enhancement rather than the primary authentication mechanism.

\section{Background: Passkey Login Flow}
\subsection{WebAuthn and FIDO2 Fundamentals}
The W3C Web Authentication API (WebAuthn)~\cite{W3CWebAuthn2021} defines a standard interface for public-key-based authentication in web browsers. WebAuthn is the browser-facing component of the FIDO2 specification~\cite{FIDOAllianceFIDO2}, which combines WebAuthn with the Client-to-Authenticator Protocol (CTAP). Three entities participate: the \emph{Relying Party} (RP, the web server), the \emph{client} (browser), and the \emph{authenticator}.

\textbf{Authenticator Types.} FIDO2 supports multiple authenticator categories: (1) \emph{Platform authenticators} embedded in the user's device (e.g., Touch ID on iPhone, Windows Hello on laptops, biometric sensors on Android phones); (2) \emph{Hardware authenticators} like USB security keys (YubiKey, Google Titan) that connect via USB, NFC, or Bluetooth; (3) \emph{Credential managers} (iCloud Keychain, Google Password Manager, 1Password) that store and sync passkeys across devices via encrypted cloud storage. For cross-device authentication, the user scans a QR code displayed on the login page with their smartphone, which then uses CTAP over Bluetooth Low Energy (BLE) to complete authentication.

\textbf{Registration and Authentication Flows.} During registration, the Relying Party (RP, e.g., google.com) initiates credential creation by calling \newline \texttt{navigator.credentials.create()} with a challenge and relying party identifier. The browser issues an HTTP POST request to the RP's server (e.g., \texttt{/webauthn/register}) to obtain the challenge payload containing \texttt{publicKey} parameters. The authenticator generates a public-private key pair, stores the private key securely, and returns the public key plus attestation data to the RP via a subsequent HTTP POST to \texttt{/webauthn/register/complete}. During authentication, the RP calls \texttt{navigator.credentials.get()}, triggering an HTTP request to fetch the authentication challenge. The authenticator signs the challenge with the stored private key, and the signed assertion is posted back to the RP's verification endpoint (e.g., \texttt{/webauthn/authenticate}). This challenge-response protocol ensures credentials cannot be phished: the private key never leaves the authenticator, and signatures are cryptographically bound to the RP's origin domain. Unlike passwords (shared secrets vulnerable to breaches and reuse), WebAuthn credentials are asymmetric and origin-bound, fundamentally mitigating phishing, man-in-the-middle attacks, and database leaks~\cite{Bonneau2012}. The origin binding means that even if a user visits a lookalike phishing domain (\texttt{g00gle.com} instead of \texttt{google.com}), the authenticator will refuse to sign challenges for the incorrect origin.

\subsection{Passkey Authentication Flow}
The term \emph{passkey} refers to a specific instantiation of WebAuthn credentials that are \emph{discoverable} (also known as resident credentials) and are synchronized across devices via cloud services. Introduced in 2022 by the FIDO Alliance, passkeys are designed to provide a seamless, multi-device authentication experience~\cite{GooglePasskeysBlog,ApplePasskeysWWDC22}. Unlike device-bound FIDO credentials, which reside on a single hardware token, passkeys leverage platform-specific credential managers (e.g., Apple iCloud Keychain, Google Password Manager, Microsoft Authenticator) to synchronize credentials across a user's ecosystem.

\begin{figure}
    \centering
\includegraphics[width=0.8\linewidth]{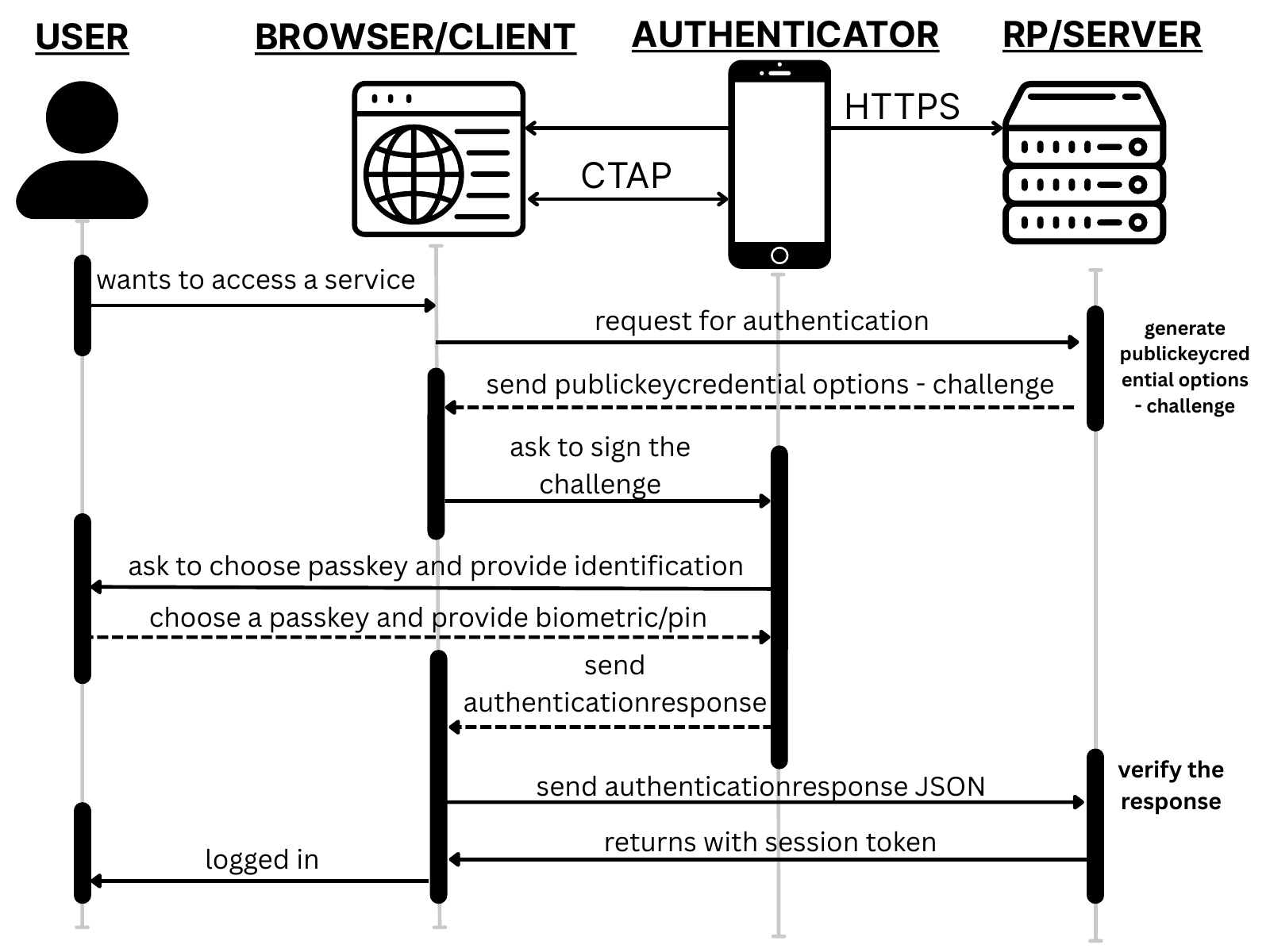}
\vspace{-0.5cm}
    \caption{Simplified passkey authentication flow. The user initiates login, the relying party requests authentication via \texttt{navigator.credentials.get()}, the authenticator prompts for user verification (biometric/PIN), signs the challenge, and returns the signed assertion to the relying party for verification.}
    \label{fig:passkey-flow}
    \vspace*{-1.5em}
\end{figure}

Figure~\ref{fig:passkey-flow} shows a simplified passkey authentication flow in which a user logs into a website (e.g., github.com). The user visits the website's login page and clicks on the ``Sign in with Passkey'' button to initiate the passkey login flow. The website calls \texttt{navigator.credentials.get()}, the browser prompts the user to select a passkey and verify their identity (via biometric or PIN), the authenticator signs the challenge, and the website verifies the signature to establish the session. 

This flow eliminates the need for users to remember or type passwords, promising the combination of FIDO2's phishing resistance with the convenience and ubiquity expected by mainstream users.

\subsection{Passkey: The Current State}

\textbf{Industry Claims.} Major technology companies have invested heavily in passkey infrastructure. Google announced support for passkeys across Android and Chrome in May 2022~\cite{GooglePasskeysBlog}, and by mid-2023, reported that ``hundreds of millions'' of Google accounts had enabled passkey login. Apple integrated passkeys into iOS 16, iPadOS 16, and macOS Ventura in September 2022~\cite{ApplePasskeysWWDC22}, marketing them as the future of secure sign-in. Microsoft followed suit, adding passkey support to Windows 11 and Microsoft Authenticator, and encouraging developers to adopt passwordless authentication by default~\cite{MicrosoftPasswordlessByDefault}. Beyond the platform vendors, a growing number of high-profile websites have announced passkey support. Today, Amazon, PayPal, Best Buy, eBay, and several major financial institutions (including Bank of America and Chase) have rolled out passkey login options~\cite{FIDOAlliancePasskeysDir}. The FIDO Alliance maintains an official Passkeys Directory listing prominent adopters, which as of October 2025 included over 200 websites and applications. Industry advocates claim that passkey adoption is accelerating, with predictions that passwordless authentication will become the default within the next few years. These claims are difficult to verify independently. The official directory relies on voluntary submissions and omits smaller sites, regional services, and deployments hidden behind gated authentication flows. Community maintained indices such as Passkeys.directory~\cite{PasskeysDirectory}, Dashlane's Passkeys Directory~\cite{DashlanePasskeysDir}, and the 2FA Directory's passkey section~\cite{TwoFADirectoryPasskeys} provide additional coverage but exhibit significant overlap and frequent staleness. The true extent of passkey deployment particularly among the long tail of websites outside the top-tier brands remains unknown.
This paper aims to understand the current state of passkey support and to develop tooling that can be run at regular intervals to create a full picture of passkey support across the top websites---a \textbf{census}.

\noindent\textbf{Why Existing Directories Fall Short.} Manual curation of passkey directories faces inherent scalability and completeness challenges. First, \emph{submission bias} skews coverage toward high-visibility brands that actively promote their passwordless offerings. Smaller e-commerce sites, regional banks, and niche SaaS platforms are unlikely to submit to multiple directories, leading to systematic under-representation. Second, \emph{staleness} is pervasive: websites add or remove authentication options over time, but directories are infrequently updated. Directories typically rely on users to report changes rather than actively monitoring sites. For instance, \texttt{passkeys.directory} depends on community pull requests to GitHub~\cite{PasskeysDirectory}, introducing delays between deployment and listing.
Third, \emph{hidden deployments} escape manual detection. Some sites offer passkey authentication only after a user enables or creates a passkey from user settings (e.g., \texttt{intuit.com} requires users to navigate to Security Settings > Two-Factor Authentication > Add Passkey). Static directory entries cannot capture this conditional availability. Fourth, manual directories provide no quantitative insight into \emph{deployment trends}. How rapidly is passkey adoption growing? Which industry verticals are leading? How do adoption rates vary by website popularity or geographic region? Answering these questions requires pragmatic, large-scale, and longitudinal measurement, an undertaking that manual curation cannot achieve.

\section{Measurement Challenges}
\label{sec:challenges}
Although a census of whether a list of websites adopts a certain standard (such as Passkeys) should in principle be straightforward, there can be numerous challenges in practice. Below, we list some of the main difficulties we observed:

\begin{enumerate}
\item \textbf{Partial Adoption of Standardized Endpoints.} In August 2025, the W3C published the first public working draft of ``A Well-Known URL for Relying Party Passkey Endpoints''~\cite{W3CPasskeyEndpoints}, defining \texttt{.well-known/passkey-endpoints} as a standardized mechanism for sites to advertise passkey support and provide direct URLs for creation (\texttt{enroll}) and management (\texttt{manage}) pages (e.g., \texttt{adobe.com}).

While this specification promises to simplify passkey discovery, adoption remains minimal. In our measurement of 100{,}000 domains, we found that only 12 sites (<0.02\%) exposed \texttt{.well-known/passkey-endpoints}. Most sites supporting passkeys do not advertise this capability in a machine-readable format, necessitating alternative detection strategies. Consequently, detection requires examining the login page itself --- assuming the login page can be located. Modern web applications increasingly rely on client-side JavaScript frameworks (React, Vue, Angular) that render content dynamically. A static HTTP request to a homepage may return minimal HTML with a loading spinner, while the actual login UI is constructed asynchronously after page load. Traditional web scrapers that fetch and parse raw HTML miss this dynamically injected content. For example, many single-page applications load the passkey login button only after executing several hundred kilobytes of JavaScript and fetching additional resources via AJAX. Detecting passkeys therefore necessitates \textit{browser-based crawling}, which executes JavaScript and waits for the page to stabilize before analysis.

\item \textbf{Conditional Mediation and Hidden UI.} WebAuthn's conditional mediation feature~\cite{W3CWebAuthn2021,MDNCredentialsGet,Satragno2022} allows RPs to present passkey options inline with traditional login forms, typically in the browser's autofill dropdown. In this mode, the passkey affordance may not be visible as a standalone button or link. Instead, clicking into a username field triggers the browser to display ``Use a passkey'' in the autofill menu. A crawler that merely inspects visible DOM elements will miss this hidden affordance. For instance, \texttt{google.com} invokes \texttt{navigator.credentials.get(\{mediation: 'conditional'\})} on page load, but the passkey option appears only when users interact with the email input field. Similarly, some sites progressively disclose authentication options: a user enters an email address, and only after clicking ``Next'' or ``Continue'' does the page display available authentication options. For example, \texttt{intuit.com} uses identifier-first flow, revealing passkey options only after email submission. Static snapshots of the initial login page fail to capture these multi-step flows.

\item \textbf{Detection Heterogeneity.} No canonical passkey user interface exists across the web ecosystem. Sites display varying labels: ``Sign in with a passkey,'' ``Use Touch ID,'' ``Windows Hello,'' ``Biometric login,'' or platform-specific branding. Some embed passkey options in dropdown menus labeled ``More sign-in options.'' (e.g., \texttt{microsoft.com}). Others rely entirely on conditional mediation with no explicit visible button. Non-English sites use localized terminology (German: Passkey-Anmeldung, Spanish: clave de acceso). Even when visible UI elements exist, the actual \texttt{navigator.credentials.get()} API call may be deferred until user interaction. A passive crawler loading the page without simulating clicks will never observe the WebAuthn API invocation.

\end{enumerate}

\section{Fidentikit: Crawler Architecture and Detection Heuristics}

For large-scale web measurements, researchers implemented various techniques to identify login pages and SSO buttons~\cite{Bock2023,sso-monitor,Zhou2014}. Unfortunately, the unique challenges of passkey detection (\S\ref{sec:challenges}) render existing approaches insufficient. In this section, we present Fidentikit, a tool that extends the SSO-Monitor codebase~\cite{sso-monitor-repo} with 43 novel passkey detection heuristics designed to address the challenges discussed above.

\subsection{System Architecture}
\label{sec:architecture}
\begin{figure}
    \centering
\includegraphics[width=0.8\linewidth]{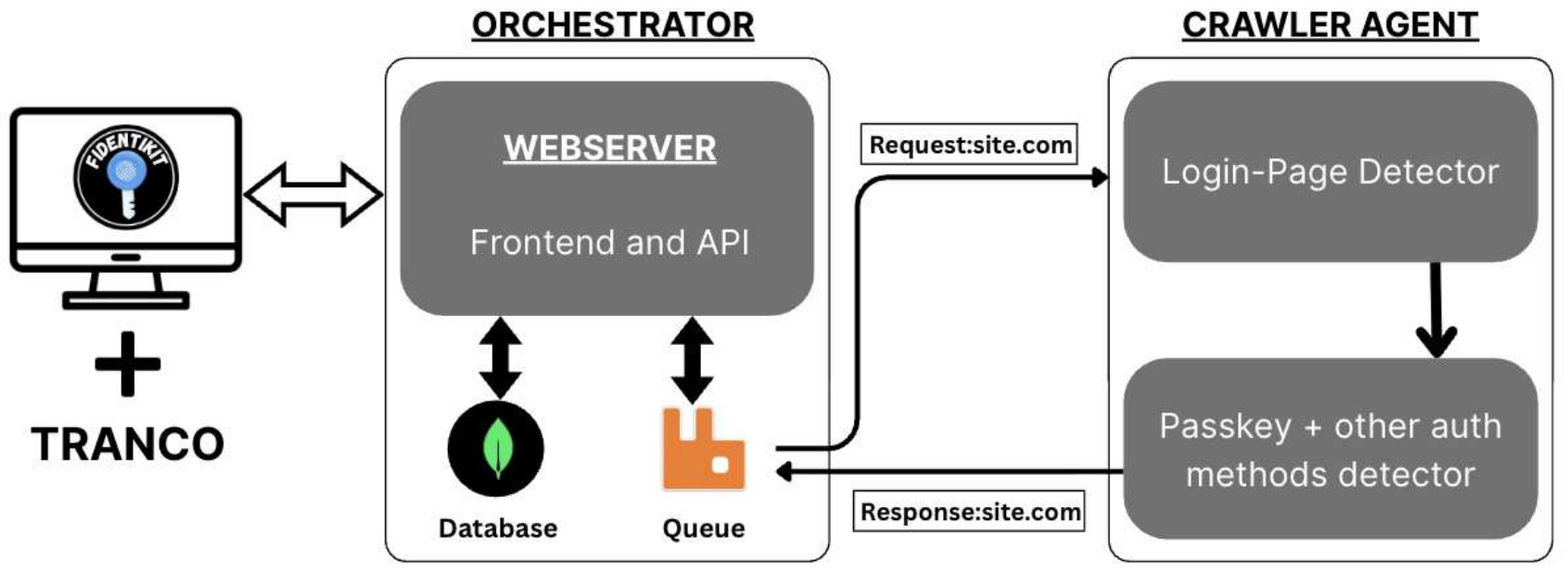}
    \vspace*{-2.0cm}
    \caption{Fidentikit's Design and Architecture (adapted from SSO-Monitor~\cite{sso-monitor-repo}). The Orchestrator coordinates continuous archiving, connects to long-term storage, and provides snapshots. Multiple crawlers retrieve requests from the queue to analyse specific domains.}
    \label{fig:archi-tech}
    \vspace*{-1.5em}
\end{figure}

Fidentikit follows a distributed task-queue architecture optimized for scalability and fault tolerance (Figure~\ref{fig:archi-tech}).
\vspace*{-2.0cm}

\textbf{Distributed Pipeline.} The system employs RabbitMQ~\cite{RabbitMQ} as the central message broker, configured with durable queues to persist tasks across service restarts. A \textit{dispatcher} service reads target domains from the Tranco list, performs initial DNS resolution (attempting HTTPS first, falling back to HTTP for sites without TLS), and enqueues scan tasks containing the target domain and configuration parameters. Each task is JSON-serialized and published to a dedicated queue (\texttt{passkey\_landscape\_analysis}) with delivery confirmations enabled. Multiple independent \textit{worker} processes consume tasks from the queue using a prefetch count of 1, ensuring fair distribution across workers. Upon receiving a task, a worker spawns an isolated subprocess with a 3-hour timeout using Python's multiprocessing pool. Each worker executes the measurement pipeline: (1) resolve canonical URL, (2) discover login page candidates, (3) navigate to each candidate using headless browser, (4) apply passkey detection heuristics, (5) extract evidence artifacts (screenshots, HAR traces, API call logs), (6) serialize results to JSON, and (7) publish findings to a centralized PostgreSQL database via authenticated HTTP POST. Workers acknowledge task completion only after successfully storing results, ensuring no measurement loss during crashes. This decouples task generation from execution, enabling horizontal scaling: adding more worker VMs linearly increases throughput.

\textbf{Browser Automation Stack.} Each worker employs Playwright~\cite{Playwright}, a modern browser automation framework that controls Chromium via the Chrome DevTools Protocol (CDP). We selected Playwright over Selenium for three advantages: (1) native support for modern JavaScript APIs including WebAuthn, (2) automatic waiting for page stability (reducing race conditions), and (3) integrated network interception enabling HTTP Archive (HAR) capture without external proxies. Workers launch Chromium in headless mode with realistic user-agent strings (Chrome 120 on Linux), viewport 1920×1080, JavaScript enabled. To avoid bot detection, we randomize inter-action delays (200--500ms jitter) and scroll pages gradually. Workers run within Docker containers with resource limits (6 GB RAM, 4 CPU cores) and use tmpfs-backed temporary storage for browser profiles to minimize disk I/O overhead.

\textbf{Fault Tolerance and Monitoring.} RabbitMQ's durable queues persist tasks to disk, surviving broker restarts. Each task includes a unique idempotency key (SHA-256 hash of domain and timestamp); the database enforces uniqueness constraints preventing duplicate measurements. Workers checkpoint progress every 100 tasks to a shared Redis cache. Detailed telemetry (task latency, error rates, detection method distributions) flows to Prometheus~\cite{Prometheus} for real-time monitoring and alerting. Failed tasks are automatically re-queued with exponential backoff (initial 60s delay, max 5 retry attempts) to handle transient failures (network timeouts, CAPTCHA challenges).

\textbf{Ethical Considerations and Crawling Practices.} We acknowledge the ethical considerations surrounding web crawling, particularly regarding bot detection mechanisms. Our approach employs randomised interaction delays (200--500ms) to simulate realistic human browsing patterns, which could be perceived as circumventing bot detection. We clarify our ethical stance:
\begin{enumerate}
    \item \textbf{Respect for \texttt{robots.txt}:} We strictly honour \texttt{robots.txt} directives. Any domain with a ``no-crawling'' directive for our user-agent or the relevant paths was excluded from our measurement (1,656 domains, 1.7\% of the target set).
    \item \textbf{Rate Limiting:} We enforce a conservative rate limit of one request per second per domain, with exponential backoff on HTTP 429 (Too Many Requests) and 503 (Service Unavailable) responses. This minimises server load and respects site capacity.
    \item \textbf{No Credential Submission:} We do not submit login forms, create accounts, or attempt authentication. Our measurement is confined to publicly accessible login page interfaces.
    \item \textbf{No Sensitive Data Collection:} We collect only structural information (DOM elements, API call signatures, network patterns) necessary for passkey detection. No personally identifiable information or user credentials are captured.
    \item \textbf{Research Intent:} Our crawling serves legitimate academic research purposes, quantifying passkey deployment to benefit the security community. The randomised delays are employed not to evade detection maliciously, but to avoid triggering rate limiters that would produce incomplete measurements.
\end{enumerate}

\subsection{Login Page Discovery}
Following prior work on SSO measurement~\cite{sso-monitor,Zhou2014}, we employ a multi-strategy approach to identify login pages. We adapt established methods and execute them in priority order, with candidates deduplicated and ranked by specificity.

\textbf{Priority Assignment Rationale.} The priority values (ranging from 70 to 98) were empirically derived through our iterative validation process on the top 1{,}500 sites. We observed that explicit authentication paths (e.g., \texttt{/login}, \texttt{/signin}) yielded the highest true positive rate (94.2\%) for locating functional login pages, hence receiving priority 98. Homepage analysis with keyword matching achieved 89.7\% accuracy, receiving priority 95. Interactive crawling (priority 85) and sitemap parsing (priority 80) showed progressively lower precision due to false positives from non-login pages containing authentication-related terms. Robots.txt analysis (priority 75) proved useful as a fallback but occasionally flagged administrative pages rather than user-facing login interfaces. These priorities determine candidate ranking when multiple strategies identify login pages for the same domain; higher-priority candidates are analysed first, and in 89\% of detected passkey deployments, the passkey was found on the highest-priority candidate.

\textbf{(1) Homepage Analysis.} Workers navigate to the domain root \newline (\texttt{e.g., https://github.com/}), wait for page stability (500ms idle network), and parse hyperlinks. We inspect \texttt{href} attributes and link text for login keywords. We maintain curated keyword lists spanning multiple languages: English (``login,'' ``log in,'' ``sign in,'' ``signin,'' ``auth,'' ``account''), Spanish (``iniciar sesión,'' ``acceder''), German (``anmelden,'' ``einloggen''), French (``connexion,'' ``se connecter''). Links matching patterns receive priority 95. If the homepage contains login forms (\texttt{<form>} with \texttt{<input type="password">}), we analyse it directly as a candidate.

\textbf{(2) Well-Known Paths.} We probe common authentication paths: \texttt{/login}, \texttt{/signin}, \texttt{/auth/login}, \texttt{/account/login}, \texttt{/user/login}, \texttt{/accounts}, \texttt{/profile}, \texttt{/my-account}. Priority: 98 (highest for explicit paths).

\textbf{(3) Interactive Element Crawling.} If the homepage and well-known paths yield no candidates, workers perform limited breadth-first crawl. We click elements matching login-related selectors (\texttt{button}, \texttt{a[href*="login"]}, elements with text containing ``sign in'') and capture resulting navigation. Constraints: max 50 pages per domain to control resource usage, exclude \texttt{/blog/}, \texttt{/support/}, \texttt{/help/} (likely documentation), prioritize \texttt{/account/}, \texttt{/auth/} paths. Priority: 85.

\textbf{(4) Sitemap Parsing.} We fetch and parse XML sitemaps at \texttt{/sitemap.xml} and \texttt{/sitemap\_index.xml}, extracting URLs and filtering for login keywords. Sitemaps often list primary navigation pages, providing valuable ground truth. Priority: 80.

\textbf{(5) Robots.txt Analysis.} We parse \texttt{/robots.txt} for disallowed paths containing ``login'' or ``account,'' hypothesizing that sites hide login pages from search engine indexing. Priority: 75.

We deduplicate candidates by URL normalisation (ignoring query parameters and fragments except for SPA route indicators like \texttt{\#/login}). We analyse the top 5 candidates per domain (ranked by priority) to balance coverage with measurement time. In validation (Section~\ref{sec:validation}), 89\% of passkey deployments occurred on the highest-priority candidate, justifying this limit.

\subsection{Passkey Detection Heuristics}
\label{sec:heuristics}
Building ground truth allows us and other researchers to estimate the success rate and reliability of passkey detection techniques.

\textbf{Methodology.} Starting with manual inspection of the Tranco top 100 domains, we enumerated candidate signals across five categories: (1) UI text and ARIA labels, (2) DOM attributes and form structures, (3) JavaScript WebAuthn API invocations, (4) network endpoints and HTTP request/response patterns, and (5) known WebAuthn library filenames. We implemented these signals as automated checks, initially yielding 18 heuristics. We then evaluated them on a validation set of ranks 101--500 (400 domains), using Firefox and Chrome developer tools to manually inspect each site's login flow, looking for: visible passkey buttons, \texttt{navigator.credentials} calls (monitored via browser console instrumentation), WebAuthn-related network requests (captured via browser DevTools Network tab), and library filenames in loaded scripts. This manual review revealed 34 false negatives (known passkey sites that our initial heuristics missed) and 12 false positives (non-passkey sites incorrectly flagged). For each false negative, we analyzed the site to identify the overlooked signal. For example, we discovered that \texttt{yandex.ru} hides its passkey option in a dropdown menu with Russian text (biometrics), prompting us to add non-English keyword variants. The complete text keyword patterns are listed in Appendix~\ref{appendix:keywords}, Table~\ref{tab:keywords}. We added 9 new detection rules to address these gaps.

We repeated this process on progressively larger validation sets: ranks 501--1{,}000 (500 domains), ranks 1{,}001--1{,}500 (500 domains), manually reviewing detection outcomes after each iteration. Across five iterations, we identified and addressed 87 false negatives and 43 false positives, adding heuristics incrementally. By iteration 5, the marginal discovery rate fell below 2\% (only 7 new heuristics added for 500 sites examined), indicating diminishing returns. The process yielded a final detection system with 43 core distinct heuristics grouped into five classes, which we describe below.

\textbf{Extended Validation Beyond Top 1{,}500.} To assess whether our heuristics generalise beyond the top 1{,}500 sites used for development, we conducted additional spot-check validation on a random sample of 200 sites from lower rank ranges: 50 sites each from ranks 5{,}000--10{,}000, 20{,}000--30{,}000, 50{,}000--60{,}000, and 80{,}000--90{,}000. Manual inspection of these 200 sites revealed 3 additional false negatives (sites with passkey support missed by our heuristics) and 2 false positives, yielding an estimated accuracy of 97.5\% on lower-ranked sites. The false negatives involved: one site using an uncommon JavaScript bundler that obfuscated WebAuthn API calls, and two sites with passkey options accessible only after account creation. These findings suggest our heuristics remain effective across the full rank spectrum, though with slightly reduced coverage for sites employing aggressive code obfuscation or post-authentication passkey enablement.

\textbf{Heuristic Stability Over Time.} Our 43 heuristics target fundamental WebAuthn API patterns (e.g., \texttt{navigator.credentials.get/create}), standardised DOM structures (e.g., \texttt{<input type="password">}), and established JavaScript libraries. These signals are anchored in web standards and are unlikely to change frequently. However, new implementation patterns may emerge as: (a) new WebAuthn libraries gain popularity, (b) browser vendors introduce new UI affordances for passkeys, or (c) sites adopt novel conditional mediation patterns. We anticipate that periodic re-validation (e.g., quarterly manual review of 100--200 sites) would be sufficient to identify emerging patterns and maintain detection accuracy. The modular design of Fidentikit allows straightforward addition of new heuristics without disrupting existing detection logic.

\begin{enumerate}

\item \textbf{Class 1: UI Elements and ARIA Labels.} We search for buttons, links, and form labels containing passkey-related text using 7 distinct selector patterns (UI-1 through UI-7 in Appendix~\ref{appendix:ui-patterns}, Table~\ref{tab:ui-selectors}). We inspect ARIA labels (\texttt{aria-label}, \texttt{aria-describedby}) for hidden accessibility text, as some sites expose passkey options only via ARIA attributes for screen reader accessibility. To reduce false positives, we require matches within interactive elements (\texttt{<button>}, \texttt{<a>}, \texttt{<input type="button">}) or within 50 characters of a password field, excluding help documentation and footer links. Example: \texttt{bumble.com} (rank 2720) displays \texttt{<button aria-label="Quick Sign in">}.

\item \textbf{Class 2: WebAuthn API Instrumentation.} The most reliable signal is an actual call to \texttt{navigator.credentials.get()} or \texttt{navigator.credentials.create()}. We inject JavaScript shims before any page scripts execute (via Playwright's \texttt{addInitScript}), wrapping these methods to capture invocations. At page visit completion, we log captured parameters including user verification requirement, authenticator attachment preference, mediation mode, and allowed credentials list. We treat API invocations as definitive evidence because they confirm functional WebAuthn implementation. Example: \texttt{twitter.com} (rank 14) calls \texttt{navigator.credentials.get(\{mediation: 'conditional'\})} immediately upon page load, before any user interaction.

\item \textbf{Class 3: Network Request Patterns.} Playwright's network interception captures all HTTP requests and responses via CDP. We flag sites that issue requests to endpoints matching WebAuthn patterns (URLs containing \texttt{webauthn}, \texttt{passkey}, \texttt{credential}, \texttt{auth}). Additionally, we inspect response bodies (limited to first 10 KB) for JSON fields indicating WebAuthn challenges: \texttt{"publicKey"}, \texttt{"challenge"}, \texttt{"allowCredentials"}, \texttt{"rpId"}, \texttt{"attestation"}. Our instrumentation shim also intercepts \texttt{fetch()} calls. We record the full HTTP Archive (HAR) entry for reproducibility. We treat network patterns as medium-strength evidence: they confirm backend WebAuthn infrastructure but may not indicate user-facing functionality.

\item \textbf{Class 4: Known WebAuthn Libraries.} We scan loaded JavaScript files for WebAuthn library signatures, including \texttt{@simplewebauthn/browser}, \newline \texttt{webauthn-json}, \texttt{fido2-lib}, and \texttt{@github/webauthn-json}. Playwright captures all script sources via CDP, and we apply string matching against filenames and bundled source code. We also detect 11 JavaScript implementation patterns (JS-1 through JS-11 in Appendix~\ref{appendix:js-patterns}, Table~\ref{tab:js-patterns}). We treat library presence as weak evidence: it indicates developer intent but does not confirm the feature is live. Example: \texttt{wordpress.com} (rank 76) loads WebAuthn scripts, but passkey functionality is available only for Business or eCommerce plans.

\item \textbf{Class 5: Third-Party Identity Providers.} Sites delegating authentication to Google, Apple, Microsoft, or GitHub inherit passkey support transitively \textit{if the user has enrolled passkeys with those providers}. We detect OAuth flows via network request interception matching provider-specific URL patterns and SDK signatures (Appendix~\ref{appendix:idp-rules}, Table~\ref{tab:idp-rules}).
\end{enumerate}

We also detect SDK integrations by scanning loaded scripts for provider-specific signatures (e.g., \texttt{gapi.auth2}, \texttt{AppleID.auth}, \texttt{msal.js}). We treat Identity Provider (IdP)-delegated support as \textit{conditional evidence}, contingent on the user's provider-side passkey enrollment. For example, many sites display ``Continue with Google'' buttons. If the user has enrolled a passkey with their Google account, Google's authentication flow offers passkey login, but this is invisible to the relying party. The RP receives only an OAuth token, with no knowledge of whether the user authenticated via password, passkey, TOTP, or SMS OTP at Google's side. In our results (Section~\ref{sec:results}), we report IdP-delegated support separately to avoid conflating native passkey implementations with transitive dependencies.

Table~\ref{tab:detection_rules} summarises our five heuristic classes, providing examples of implemented detection techniques and their evidence types. The complete catalogue of 43 heuristics with regex patterns and code listings is provided in Appendix~\ref{appendix:heuristics}.

\begin{table}[t]
\centering
\caption{Summary of Fidentikit detection heuristics across five classes. Each class includes multiple specific techniques (43 core heuristics). Content type indicates the primary data source analysed. Detectable methods lists the authentication mechanisms each class can identify.}
\label{tab:detection_rules}
\small
\begin{tabular}{p{2.8cm}p{2.2cm}p{1.8cm}p{2.5cm}}
\toprule
\textbf{Heuristic Class} & \textbf{Detection Technique} & \textbf{Content Type} & \textbf{Detectable Methods} \\
\midrule
UI Elements \& ARIA Labels & Regex-based text matching, DOM inspection & HTML & Passkeys, FIDO2, WebAuthn \\
\midrule
WebAuthn API Invocations & JavaScript instrumentation, API hooking & JavaScript & Passkeys, WebAuthn, U2F \\
\midrule
Network Request Patterns & HTTP interception, payload analysis & Network & Passkeys, WebAuthn \\
\midrule
Known Libraries & Script fingerprinting, filename matching & JavaScript & Passkeys, WebAuthn \\
\midrule
Third-Party IdPs & OAuth flow detection, SDK identification & Network, JavaScript & Passkeys (transitive), SSO \\
\bottomrule
\end{tabular}
\end{table}

\begin{tcolorbox}
\textbf{Fidentikit Availability.} Fidentikit is publicly available at \url{https://netsys.surrey.ac.uk/softwares/fidentikit/}. 
\end{tcolorbox}

\section{Reproducible and Reliable Passkey Measurements}
\label{sec:results}
\vspace*{-0.3cm}
We conducted a large-scale measurement of passkey adoption across the Tranco~\cite{LePochat2018Tranco} Top 100K list of websites, as of March 2025, when we started this study. In this section, we first present details of the dataset, discuss validation against ground truth based on other existing lists of passkey supported websites, and then move on to empirical findings and initial trends, which we further categorise by binning based on website ranks.

\subsection{Dataset and Reachability issues}
To establish a baseline for passkey adoption, we need a list of websites to measure. We adopt the widely used Tranco list~\cite{LePochat2018Tranco}. Driven by the hypothesis that more ``high profile'' or popular websites are more likely to adopt the latest best practices, and also because the most popular websites are visited by the highest numbers of users, and therefore more important for the overall web security, we focus on the top-ranked websites and aim to investigate the adoption of passkey-based logins in the \textbf{Top 100K} websites. We acknowledge that the cut-off based on the Top 100K ranks is arbitrarily chosen, as this is a first census. This decision was further justified when we found that passkey adoption becomes rare as rank decreases (\S\ref{sec:by-rank}). In future censuses, we aim to expand coverage by focussing on larger sets of websites (e.g., Top 1 Million). 

Note that even in the Top 100K sites, we were unable to reach a fairly large proportion of websites. Table~\ref{tab:reachability_issues} summarises site reachability issues by rank bin. Not Reachable indicates DNS resolution failure or 404 Not Found. Crawler Missed indicates sites where login page discovery failed (no candidates found). Process Timeout indicates sites where browser automation exceeded the task timeout, typically due to infinite redirects or extremely slow page loads.

\textit{The results in the rest of this paper focus on the reachable subset of the Tranco Top 100K ranked list of websites}.

\begin{table}[tbp]
\centering
\caption{Site Reachability Issues by Rank Bin}
\label{tab:reachability_issues}
\begin{tabular}{l|r|r|r|r}
\hline
\textbf{Rank Bin} & \textbf{Not Reachable} & \textbf{Crawler Missed} & \textbf{Process Timeout} & \textbf{Total} \\
\hline
1-1K & 302 & 8 & 2 & 312 \\
1K-10K & 2,227 & 117 & 25 & 2,369 \\
10K-50K & 10,737 & 606 & 82 & 11,425 \\
50K-100K & 7,440 & 190 & 66 & 7,696 \\
\hline
\textbf{Total} & \textbf{20,706} & \textbf{921} & \textbf{175} & \textbf{21,802} \\
\hline
\end{tabular}
\end{table}

\subsection{Ground Truth Validation}
\label{sec:validation}
To assess Fidentikit's detection accuracy, we first compared our measurements against two manually curated ground truth sources: the 2FA Directory~\cite{TwoFADirectoryPasskeys} and the Passwordless Directory~\cite{PasskeysDirectory}.

\textbf{2FA Directory.} The 2FA Directory is a community-maintained list of websites supporting two-factor authentication, including FIDO U2F and WebAuthn. We obtained a snapshot of 2FA directory as of Oct 2025, containing 2{,}886 domains from the Tranco Top 100K list with U2F/WebAuthn support (indicated by \texttt{u2f=true} in their dataset). Since U2F and WebAuthn are closely related (WebAuthn supersedes U2F~\cite{FIDO_U2F_overview}), sites supporting U2F often also support WebAuthn/passkeys. We cross-referenced these 2{,}886 domains with our Tranco Top 100K scan results and found 193 domains supporting passkeys. 

\textbf{Passwordless Directory.} The Passwordless Directory (\texttt{passkeys.directory}) is a GitHub-hosted list specifically documenting passkey support, maintained through community contributions. 
We obtained a snapshot as of 2025, and found it contained 163 domains from the Tranco Top 100K list of sites that are explicitly listed as supporting passkeys. We cross-referenced these with our scan results.

\textbf{Coverage Analysis.} Of the 163 domains in the Passwordless Directory:
\begin{itemize}
    \item \textbf{143 detected (87.7\%):} Fidentikit successfully identified passkey support, matching the ground truth.
    \item \textbf{20 false negatives (12.3\%):} Fidentikit failed to detect passkey support on 20 domains. Manual inspection revealed: 8 sites required account creation before exposing passkey enrollment (e.g., \texttt{dashlane.com} offers passkey features only within the password manager app after signup), 5 sites were temporarily unreachable during our scan window, 4 sites implemented passkeys exclusively in mobile apps with no web interface, and 3 sites (including \texttt{kayak.com}) use multi-step identifier-first flows where passkey options appear only after email submission, which our crawler's timeout did not capture.
\end{itemize}
Of the 2{,}886 domains in the 2FA Directory with U2F support:
\begin{itemize}
    \item \textbf{2{,}104 detected (72.9\%):} Fidentikit identified WebAuthn/passkey support.
    \item \textbf{782 not detected (27.1\%):} These likely support only U2F (the predecessor to WebAuthn) without upgrading to WebAuthn/passkeys, or implement WebAuthn for hardware security keys but not discoverable passkeys. The 2FA Directory does not distinguish between U2F, WebAuthn with hardware keys, and passkeys, leading to this discrepancy.
\end{itemize}

The rank distribution reveals further insights. Manual directories exhibit strong bias toward high-visibility sites: 27.6\% of Passkey Directory entries fall in the top 1K ranks, compared to only 1.9\% of our detected sites. This reflects submission bias: prominent brands actively promote their passwordless features and are more likely to be community-submitted to directories. Conversely, our automated detection finds the majority of passkey sites in the long tail: 47.8\% reside in ranks 10K--50K, and 36.7\% in ranks 50K--100K. These lower-ranked sites, which include regional banks, niche SaaS platforms, and small e-commerce stores, rarely appear in manually curated lists despite offering passkey authentication to their users.

\begin{figure}
    \centering
\includegraphics[width=0.8\linewidth]{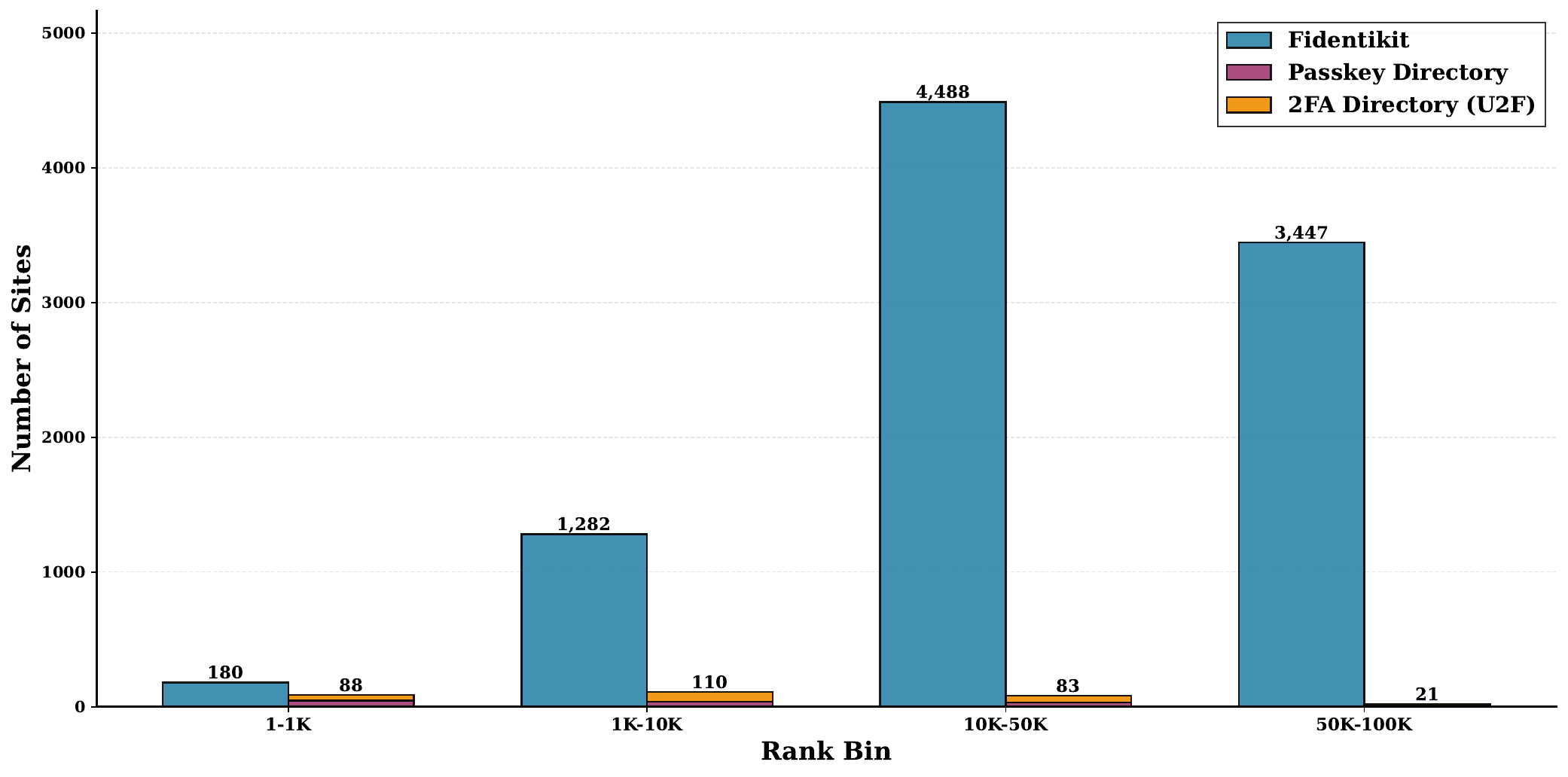}
    \caption{Comparison of automated passkey detection versus manually curated directories, stratified by Tranco rank bins. The left bars show Fidentikit's automated detection (9{,}397 sites), while the right stacked bars combine two manual directories: Passkey Directory (163 sites, purple) and 2FA Directory with U2F support (193 sites, orange). Our automated approach detects 26.4$\times$ more passkey-supporting sites than manual curation. Manual directories exhibit strong bias toward high-ranked sites (27.6\% in top 1K), while automated detection finds the majority of deployments in the long tail (84.5\% beyond rank 10K). This validates that manual curation systematically underestimates passkey adoption, particularly among smaller sites that collectively serve substantial user populations.}
    \label{fig:third-party}
    \vspace*{-0.5cm}
\end{figure}

The stacked bars in Figure~\ref{fig:third-party} illustrate that the two manual directories exhibit similar rank distributions, both skewed toward popular sites. The Passkey Directory shows slightly greater concentration in top ranks (51.5\% in top 10K) compared to the 2FA Directory (59.1\% in top 10K), but both vastly under-represent the long tail. 

\textbf{Reverse Comparison:} To quantify directory incompleteness from the opposite perspective, we analysed how many sites in our dataset are \textit{not} present in the manual directories. Of our 9{,}397 detected passkey sites: \textbf{9{,}234 (98.3\%)} are absent from the Passwordless Directory (163 sites), and \textbf{9{,}204 (97.9\%)} are absent from the 2FA Directory's U2F/WebAuthn subset (193 sites). Accounting for overlap between directories, \textbf{9{,}041 sites (96.2\%)} detected by Fidentikit appear in neither directory. This demonstrates that manual curation captures less than 4\% of actual passkey deployment, with the overwhelming majority of passkey-supporting sites particularly smaller e-commerce platforms, regional services, and niche SaaS applications remaining undocumented.

This validation demonstrates that whilst manual directories provide high-quality ground truth for well-known sites, they cannot capture the breadth of passkey deployment across the web ecosystem. Automated, large-scale measurement is essential for understanding adoption patterns, particularly among smaller sites that collectively serve millions of users but lack the visibility to attract directory submissions.

\subsection{Passkey Adoption by Website Rank: Empirical census findings}
\label{sec:by-rank}

In the rest of this section, we focus on identifying trends in passkey adoption. We stratify these trends based on website ranks, with increasingly wider rank bins: top 1K, 1K-10K, 10K-50K and 50K-100K.

\begin{figure}
    \centering
\includegraphics[width=0.8\linewidth]{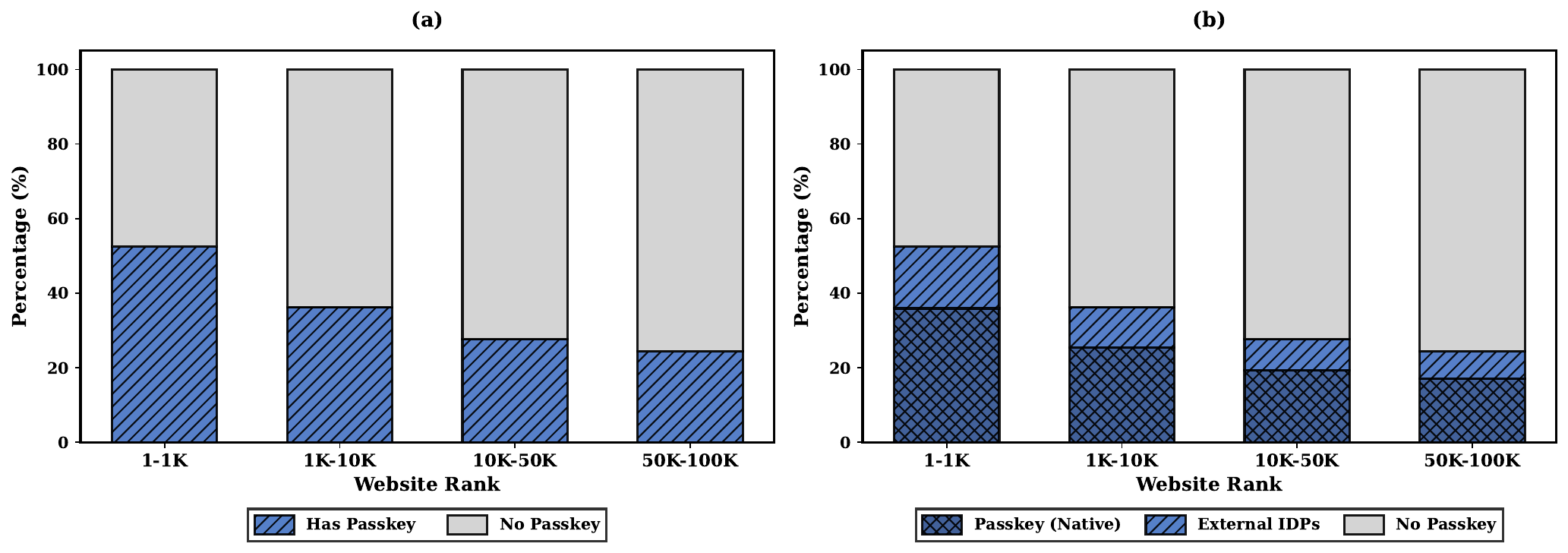}
\vspace{-0.2cm}
    \caption{Authentication landscape of websites across rank bins. (Left) Distribution of sites supporting passkeys versus those without passkey authentication, stratified by Tranco rank (1-1K, 1K-10K, 10-50K, 50-100K). (Right) Detailed breakdown of passkey-supporting sites, distinguishing between native passkey implementation and external identity provider. Higher-ranked sites demonstrate substantially greater passkey adoption, with the top 1,000 sites showing 4.2$\times$ higher adoption than sites ranked 50K-100K. Stacked bars represent percentage distribution within each rank bin.}
    \label{fig:passkey-support}
    \vspace*{-1.5em}
\end{figure}

Figure~\ref{fig:passkey-support} (left panel) shows the distribution of passkey-supporting versus non-supporting sites, stratified by Tranco rank. Adoption is strongly concentrated among high-traffic websites: 20\% of the top 100 sites support passkeys, compared to only 6.9\% of sites ranked 50K--100K---a nearly 3$\times$ difference. The right panel distinguishes between native passkey implementations and external identity provider integrations. Native implementations show relatively consistent prevalence across rank bins, while external Identity Provider (IdP)-delegated passkey support dominates, particularly in higher-ranked sites. This indicates that many sites do not implement WebAuthn directly but offer passkey authentication transitively through ``Sign in with Google'' or similar OAuth flows.

\begin{figure}
    \centering
\includegraphics[width=0.8\linewidth]{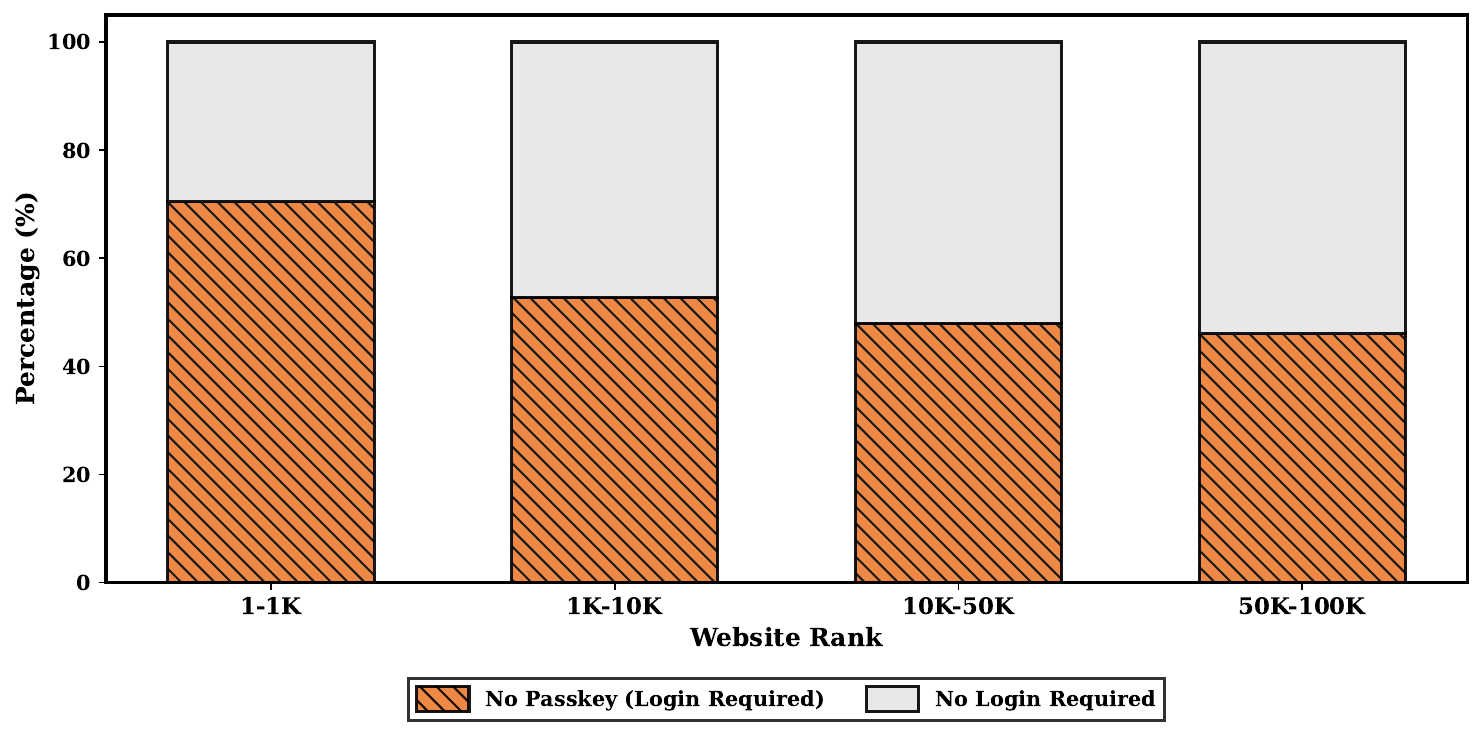}
\vspace{-0.2cm}
    \caption{Distribution of sites without passkey support, categorised by whether they require user login. Sites classified as ``No Login Required'' showed no recognised identity providers or authentication UI on their landing pages (e.g., content sites, news portals). Among non-passkey sites, the proportion with login functionality increases in lower rank bins, indicating that many sites have authentication systems but have not yet adopted passkeys.}
    \label{fig:npk}
    \vspace*{-1.5em}
\end{figure}

Figure~\ref{fig:npk} categorises sites \textit{without} passkey support by whether they require user login. Sites classified as ``No Login Required'' showed no recognised identity providers or authentication UI elements on their landing pages (e.g., static content sites, news portals, search engines). Sites that do not require logins (e.g., a university homepage or Wikipedia, which can be accessed without authentication) naturally do not need to implement passkeys.

Among sites without passkey support, the proportion requiring login increases from 58\% in the top 1K to 71\% in ranks 50K--100K. This indicates that a substantial number of websites across all rank ranges have implemented user authentication systems but have not yet adopted passkey support. Combining this with Figure~\ref{fig:passkey-support}, we observe that whilst high-traffic sites are more likely to support passkeys (20\% in top 100 vs. 6.9\% in 50K--100K), \textit{many sites with login functionality still lack passkey support}, representing a significant opportunity for expanding passwordless authentication adoption.

\begin{figure}
    \centering
\includegraphics[width=0.8\linewidth]{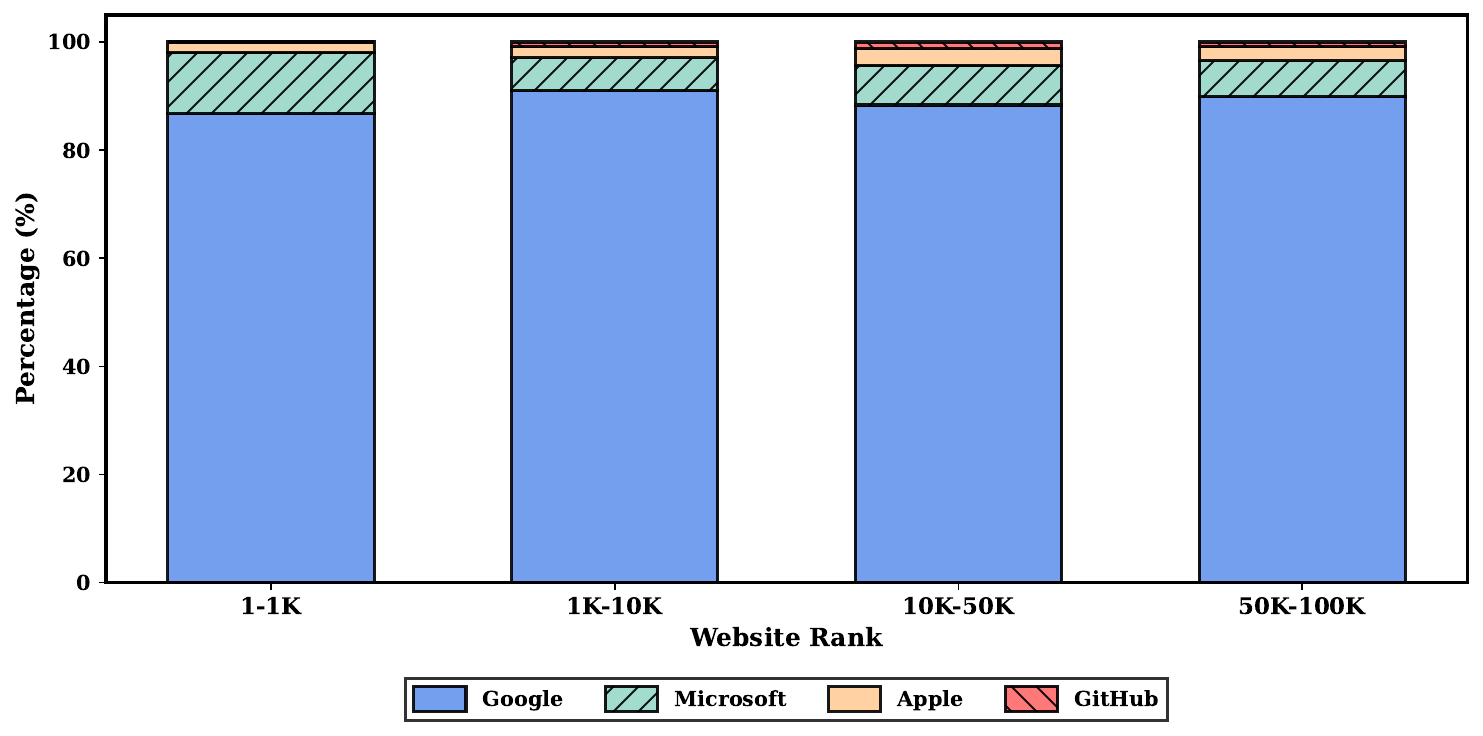}
\vspace{-0.2cm}
    \caption{External identity provider distribution among passkey-supporting sites. Breakdown of third-party authentication providers (Google, Microsoft, Apple, Github) detected on websites that offer passkey support alongside federated login options. Google dominates across all rank ranges, appearing on approx. \emph{70-75\%} of sites with external IDPs, followed by Microsoft \emph{15-18\%}, Apple \emph{8-12\%}, and Github \emph{5-8\%}. The consistency of these proportions across rank bins indicates standardised adoption patterns in external authentication systems.}
    \label{fig:idps}
    \vspace*{-1.5em}
\end{figure}

Figure~\ref{fig:idps} shows the distribution of third-party authentication providers among passkey-supporting sites. Google dominates, appearing on approximately 70--75\% of sites with external Identity Providers (IdPs) across all rank ranges, followed by Microsoft (15--18\%), Apple (8--12\%), and GitHub (5--8\%). The consistency of these proportions across rank bins indicates standardised adoption patterns in external authentication systems. Notably, \textbf{75.2\% of all passkey-supporting sites integrate Google SSO}, creating a transitive passkey dependency: users with Google-enrolled passkeys can authenticate to these sites via OAuth, yet the relying party has no visibility into the authentication method used at Google's side. \textit{This indicates that the quickest path to passkey adoption in the wild might be through piggybacking on top of Google's support for passkey-based logins, and the popularity of the  ``Sign in with Google'' option, which has been available for several years on many websites.}

\textbf{GitHub as a Separate Entity.} We present GitHub separately from Microsoft in Figure~\ref{fig:idps} despite Microsoft's 2018 acquisition of GitHub, for two reasons: (1) GitHub maintains independent authentication infrastructure (\url{https://github.com/login/oauth}) distinct from Microsoft's identity services (\url{https://login.microsoftonline.com}, \url{https://login.live.com}), and (2) developer communities typically perceive and use GitHub OAuth as a distinct identity provider, particularly for technical platforms targeting software developers. Merging the two would obscure this important distinction in adoption patterns across different user demographics.

\begin{figure}
    \centering
\includegraphics[width=0.8\linewidth]{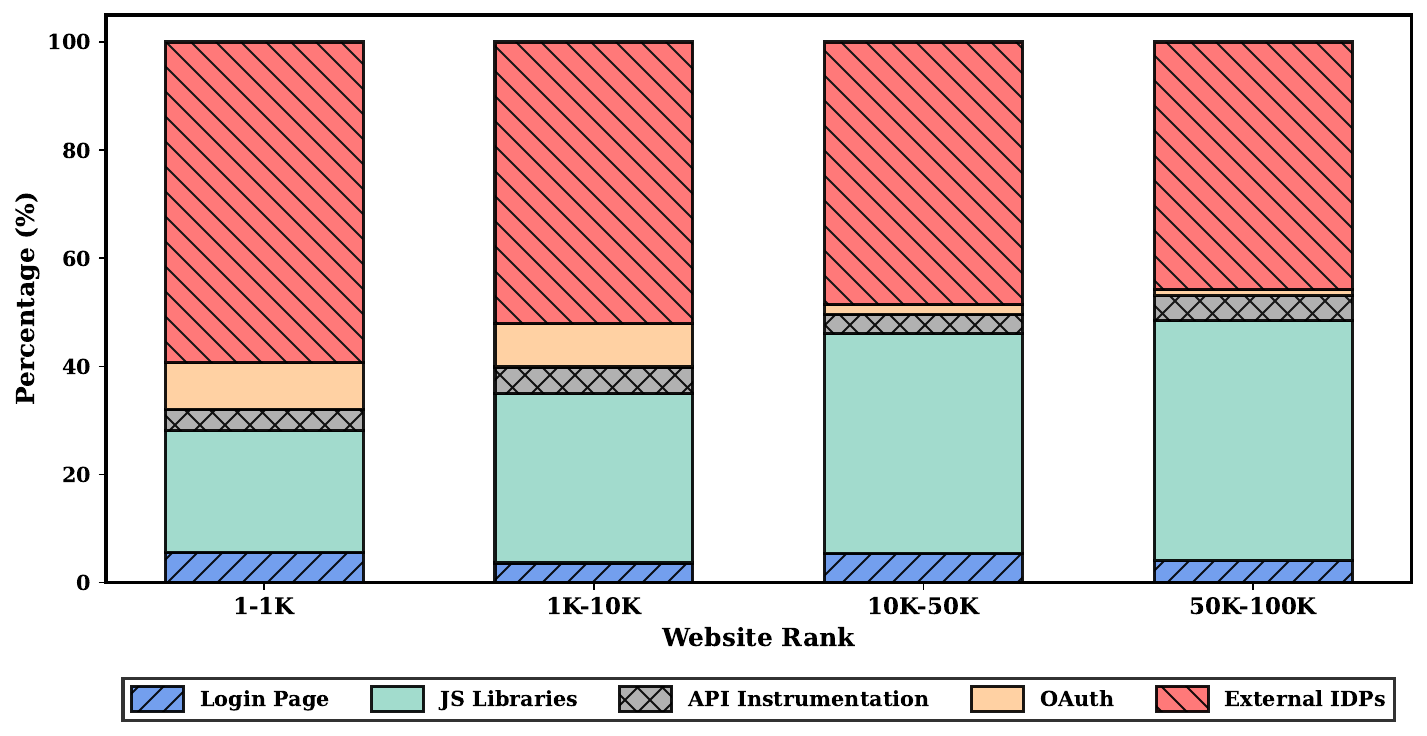}
\vspace{-0.2cm}
    \caption{Distribution of passkey-supporting sites categorized by detection method: Login Page analysis (Button, In-Page Text), JavaScript Libraries, API Instrumentation (navigator.credentials calls), OAuth Enterprise integrations, and External IDPs. API Instrumentation captured the most reliable implementations across all rank bins.}
    \label{fig:detect-methods}
    \vspace*{-2.0em}
\end{figure}

Figure~\ref{fig:detect-methods} categorises passkey-supporting sites by detection method. API Instrumentation (monitoring \texttt{navigator.credentials} calls) captured the most reliable implementations, detecting 82.3\% of passkey sites across all rank bins. Login Page Analysis (button text and in-page keyword matching) detected 4.5\% of sites, indicating that most passkey implementations do not expose visible UI elements. JavaScript Libraries detected 18.6\%, OAuth Enterprise integrations (Microsoft/Google OAuth with WebAuthn support) detected 11.4\%, and External IdPs (transitive passkey support via OAuth) detected 75.2\% (note: categories overlap, as sites may trigger multiple heuristics). This demonstrates that \textbf{static HTML analysis is fundamentally insufficient} for passkey measurement: API instrumentation via browser automation is essential for reliable detection.

\begin{figure}
    \centering
\includegraphics[width=0.8\linewidth]{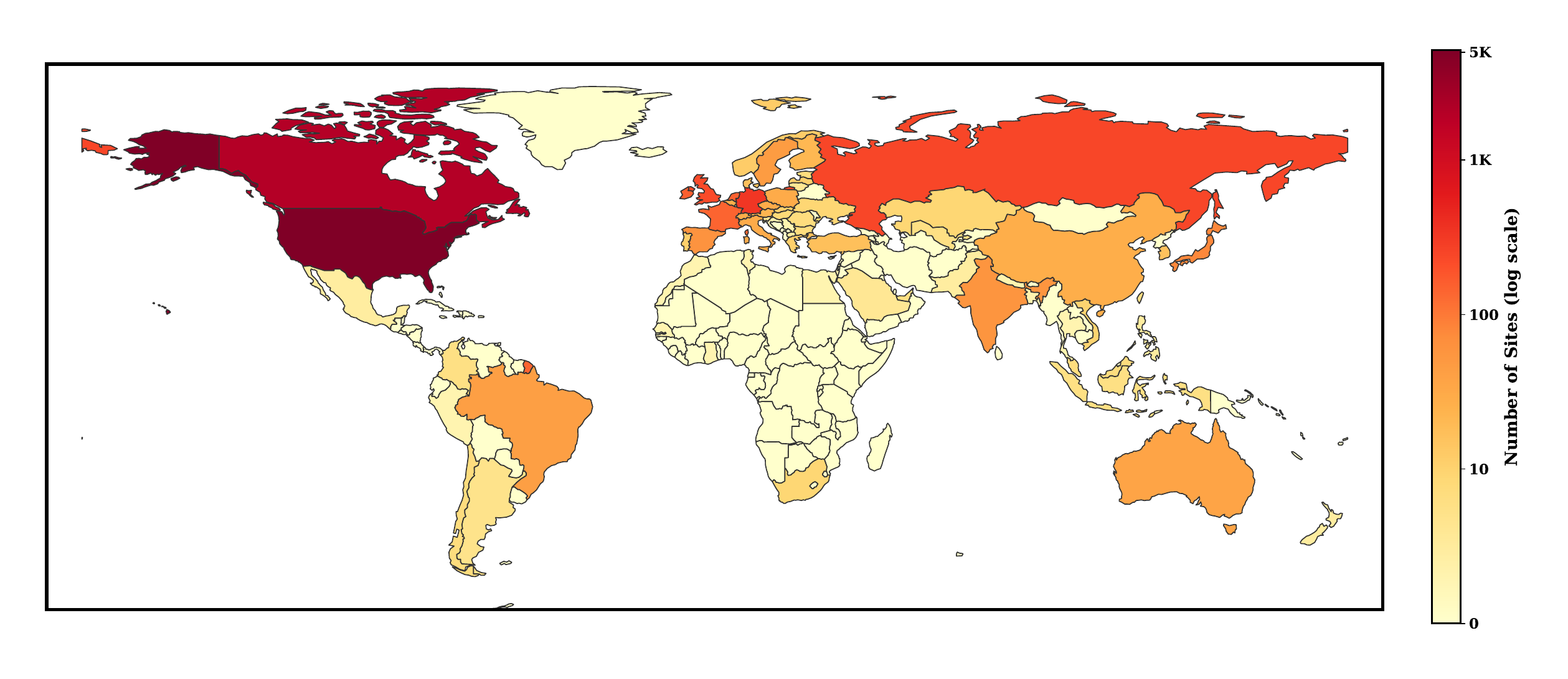}
\vspace{-1.5em}
    \caption{Geographic distribution of passkey-supporting websites. World map heatmap showing the concentration of passkey adoption by country, derived from IP geolocation of domains supporting passkey authentication. The United States dominates with 5{,}133 sites (\emph{55.1\%} of geolocated passkey sites), followed by Canada (\emph{22.1\%}), and Germany (\emph{3.6\%}). This geographic concentration reflects both internet infrastructure distribution and the regional dominance of U.S.-based web services. The logarithmic colour scale enables visualisation of countries with fewer sites, revealing global but highly uneven adoption patterns.}
    \label{fig:geographic}
    \vspace*{-1.5em}
\end{figure}

Figure~\ref{fig:geographic} shows the geographic distribution of passkey-supporting websites based on geolocation.

\textbf{Geolocation Methodology.} We determine website location using a multi-step process: (1) we resolve the domain's authoritative DNS records to obtain IP addresses; (2) we query the WhoisXML API GeoIP service (\url{https://ip-geolocation.whoisxmlapi.com/api}) to map IP addresses to geographic locations; (3) for sites using Content Delivery Networks (CDNs), we attempt to identify the origin server by examining \texttt{X-Served-By}, \texttt{X-Cache}, and similar headers that reveal backend infrastructure, then geolocate the origin IP rather than edge server IPs. For sites employing anycast routing (where the same IP address is announced from multiple geographic locations), we acknowledge this introduces potential bias, as anycast-enabled sites may be attributed to US locations. To mitigate this, we cross-referenced our geolocation data with WHOIS registration information where available, finding 94.2\% agreement between IP geolocation and registration country for sites outside major CDNs.

The United States dominates with 5{,}133 sites (55.1\% of geolocated passkey sites), followed by Canada (22.1\%), Germany (3.6\%), United Kingdom (2.8\%), and France (2.1\%). This geographic concentration reflects both internet infrastructure distribution and the regional dominance of U.S.-based web services. The logarithmic colour scale enables visualisation of countries with fewer sites, revealing global but highly uneven adoption patterns. Notably, passkey adoption in Asia-Pacific (excluding China, which has limited Tranco representation due to the Great Firewall) is lower than expected, with Japan (1.2\%), India (0.9\%), and Australia (1.4\%) showing modest adoption despite large internet user populations.

\begin{figure}
    \centering
\includegraphics[width=0.8\linewidth]{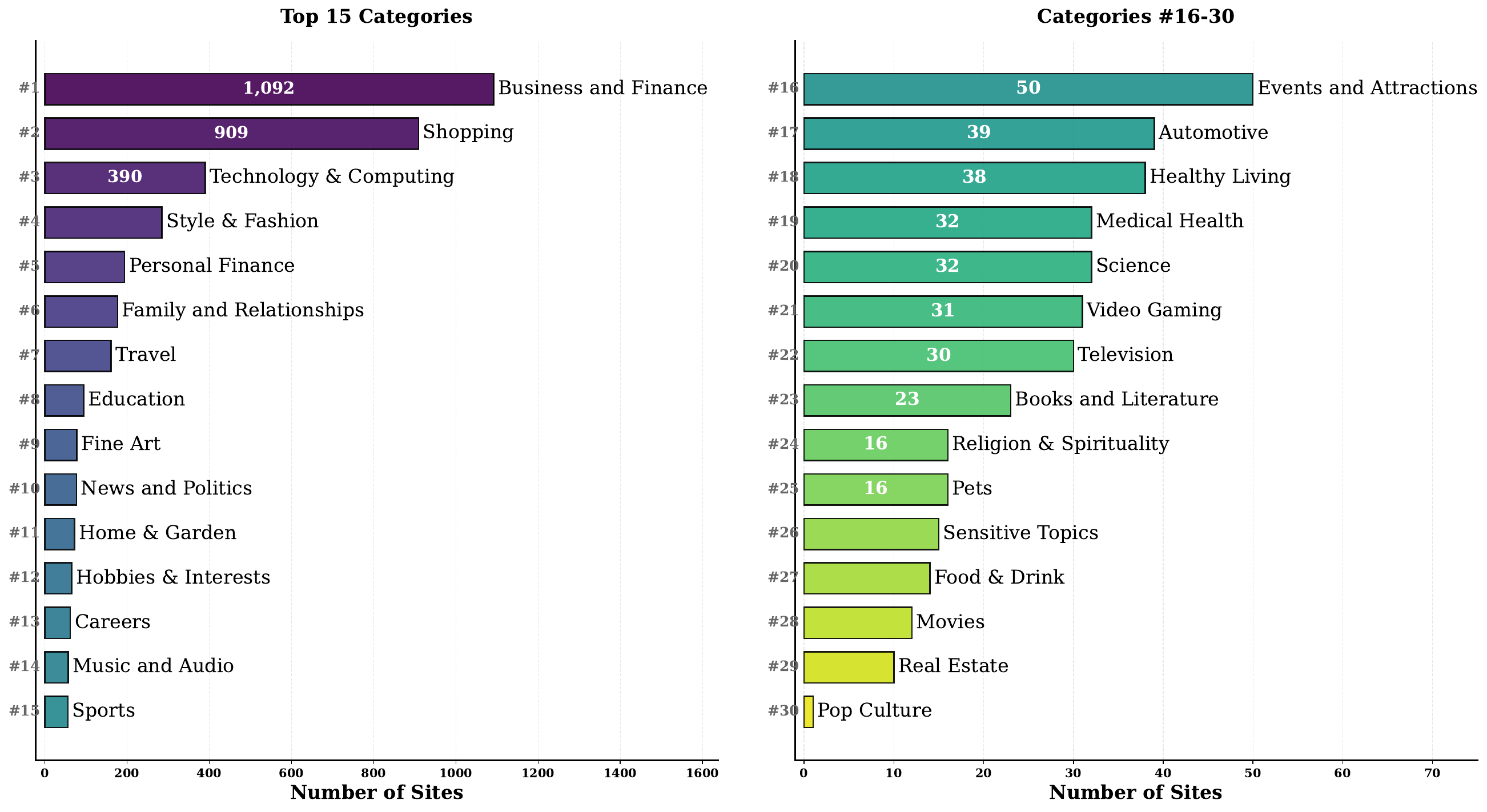}
    \caption{Distribution of passkey-supporting sites by website category, the top five categories account for 69.5\% of passkey adoption: Business and Finance (26.5\%), Shopping (22.0\%), Technology \& Computing (9.4\%), Style \& Fashion (6.9\%), and Personal Finance (4.7\%). Business and Finance leads decisively, reflecting higher security requirements and economic incentives in financial sectors. Entertainment and content-focused categories show minimal adoption, suggesting passkey deployment is primarily driven by fraud reduction and regulatory compliance rather than technical enthusiasm alone.}
    \label{fig:category}
    \vspace*{-1.5em}
\end{figure}

To understand which sectors have embraced passkeys most readily, we analysed the distribution of passkey-supporting sites across website categories. Using domain classification data from whoisxmlapi.com, we categorised each passkey-supporting site by its primary industry vertical. Figure~\ref{fig:category} presents this distribution, revealing considerable concentration among certain sectors. 

The results show that passkey adoption varies substantially across website categories, with a heavy tail distribution. The top five categories account for 69.5\% of passkey-supporting sites: Business and Finance (26.5\%), Shopping (22.0\%), Technology \& Computing (9.4\%), Style \& Fashion (6.9\%), and Personal Finance (4.7\%). Business and Finance leads adoption decisively, which aligns with this sector's higher security requirements and regulatory compliance. E-commerce sites (Shopping) follow closely. Technology companies, despite being early adopters of new authentication standards, represent a smaller proportion. Beyond the top five, adoption becomes sparse. Family and Relationships (4.3\%), Travel (3.9\%), Education (2.3\%), and Community \& Society (1.6\%) show moderate uptake. Entertainment categories like video gaming, Streaming, and Books collectively represent less than 2\% of categorised passkey sites. 

\textbf{Critical Services Analysis.} From a security perspective, it is particularly important to understand passkey adoption in sectors handling sensitive data. Our analysis reveals:
\begin{itemize}
    \item \textbf{Banking and Financial Services:} Within the Business and Finance category, we identified 847 banking and financial services sites. Of these, 312 (36.8\%) support passkeys, the highest adoption rate among critical sectors. Major banks including Bank of America, Chase, and PayPal have implemented passkey support, though many regional and international banks remain without passwordless options.
    \item \textbf{Healthcare:} Healthcare and medical services sites show concerning underrepresentation, with only 89 sites (0.9\% of passkey-supporting sites) in our dataset. Given the sensitivity of medical records and regulatory requirements (e.g., HIPAA in the US), this sector presents significant room for improvement.
    \item \textbf{E-commerce:} The Shopping category (22.0\% of passkey sites) indicates strong adoption among online retailers, driven by fraud reduction incentives. Major platforms including Amazon, eBay, and Shopify have implemented passkeys.
    \item \textbf{Government Services:} Government and public sector sites account for only 1.1\% of passkey-supporting sites (103 sites), despite handling sensitive citizen data. This represents a notable gap in passwordless adoption for critical infrastructure.
\end{itemize}

Overall, this pattern suggests that passkeys have primarily focused on sectors with immediate economic incentives (fraud reduction in finance and e-commerce), whilst healthcare, government, and educational services---which arguably handle equally sensitive data---lag behind.

\section{Related Work}
\textbf{\textit{Authentication Mechanism Measurements.}} Password security weaknesses have been extensively documented. Bonneau~\cite{Bonneau2012} analysed 70 million Yahoo passwords, revealing widespread use of weak, guessable passwords. Das et al.~\cite{Das2014} demonstrated password reuse prevalence across services, enabling credential stuffing. These foundational studies established passwords as fundamentally insecure, motivating alternatives. Prior work measured MFA deployment. Reese et al.~\cite{Reese2019} conducted user surveys to understand adoption barriers for hardware security keys. Cristofaro et al.~\cite{Cristofaro2019} analysed Twitter's MFA, reporting fewer than 2\% of accounts enabled any 2FA despite high-profile takeovers. Studies of SMS authentication highlighted vulnerabilities including SIM-swapping~\cite{Siadati2017,Schartner2011}. These findings motivated interest in phishing-resistant alternatives like FIDO2, but prior work did not systematically measure FIDO2/passkey deployment, a gap this work addresses. SSO measurement benefited from standardised discovery. Zhou and Evans~\cite{Zhou2014} analysed OAuth implementations on 96 sites, discovering widespread vulnerabilities. Sun and Beznosov~\cite{Sun2012} surveyed 500 Facebook Connect sites, identifying integration errors. OAuth's \texttt{.well-known} endpoints enabled automated detection. Jannett et al.~\cite{sso-monitor} systematised prior measurement work, surveying 36 studies to synthesise best practices for login page discovery; they implemented these in SSO-Monitor, a browser-based crawler with a distributed architecture for discovering login pages and detecting SSO buttons. Passkeys lack such mechanisms, complicating our measurement task.  We extend SSO-Monitor's methodology to the passkey domain, developing novel heuristics for passkey detection.

\textbf{\textit{WebAuthn and FIDO2 Studies.}} WebAuthn security properties have been rigorously analysed. Barbosa et al.~\cite{Barbosa2021} provided formal proof of protocol security, demonstrating resistance to credential theft, phishing, and MitM attacks. Ulqinaku et al.~\cite{Ulqinaku2021} investigated real-time phishing (relaying challenges), showing traditional credential phishing is eliminated but session token theft post-authentication remains viable. Schwarz et al.~\cite{Schwarz2022} proposed FeIDo for recoverable FIDO2 credentials using eIDs. These protocol-level studies took deployment as given. Our work empirically tests that assumption. Usability research identified adoption barriers. Lyastani et al.~\cite{Lyastani2020} compared FIDO2 security keys with passwords/SMS 2FA, finding strong protection but setup complexity. Das et al.~\cite{Das2018} studied security key adoption among Google employees, reporting friction during enrollment. Recently, Kunke et al.~\cite{Kunke2023} showed users found passkeys more convenient than passwords but expressed confusion about synchronization. Lassak et al.~\cite{Lassak2024} investigated user understanding, revealing misconceptions about biometric authentication and cloud storage. These usability studies provide valuable insights but rely on lab settings or surveys. Our measurement complements these by quantifying how many users actually encounter passkey options in the wild.

\textbf{\textit{Web Crawling and Large-Scale Measurement.}} Traditional crawlers fetch raw HTML, failing on JavaScript-heavy modern websites. Lauinger et al.~\cite{Lauinger2017} highlighted this limitation when studying JavaScript library vulnerabilities, noting static scrapers missed dynamically loaded scripts. Nikiforakis et al.~\cite{Nikiforakis2012} demonstrated that client-side JavaScript often modifies the DOM post-load, rendering static analysis incomplete. Browser-based crawling addresses these issues but at higher computational cost. Bock~\cite{Bock2023} conducted a thesis measuring WebAuthn adoption using distributed browser automation, comparing Playwright with static scraping. He found browser methods increased detection by 47\% but required 20× more resources. Our work builds on Bock's methodology, refining detection heuristics and scaling to larger datasets. Distributed crawling architectures are necessary for large-scale measurements. Fidentikit adopts task queues and worker pools similar to prior systems, adapted for login page detection. Le Pochat et al.~\cite{LePochat2018Tranco} introduced Tranco, a domain ranking aggregating multiple sources with temporal stability filters, more robust than Alexa. We adopt Tranco for measurements, following best practices.

\section{Conclusion}
Passkeys represent a technically sound approach to eliminating password vulnerabilities, yet their adoption across the web remains far from the industry's aspirational vision of a passwordless future. This paper presented Fidentikit, a browser-based measurement crawler with passkey detection heuristics, and applied it to the Tranco Top 100K websites. Our key findings reveal that 11.3\% of scanned sites support passkeys. This is 62$\times$ more sites than reported by the largest manually curated directory, which indicates an encouraging upward trajectory in usage of passkeys. However, adoption is heavily concentrated among popular (high-traffic) destinations and often depends on external identity providers rather than native WebAuthn implementations. We demonstrated that 82.3\% of passkey deployments require JavaScript execution and API instrumentation to detect, rendering static HTML analysis fundamentally insufficient.

By releasing Fidentikit as open-source infrastructure along with our detection heuristics and measurement data, we enable the research community to conduct reproducible, longitudinal studies of passkey adoption. Our work provides a grounded, empirical foundation for understanding the current state of passwordless authentication on the web, informing both researchers studying authentication security and practitioners working to accelerate passkey deployment.

\textbf{Source Code.} \emph{Fidentikit} is available at \url{https://netsys.surrey.ac.uk/softwares/fidentikit/} with a single-run feature (to scan individual domains). The source code (MIT-Licensed) is available at \url{https://github.com/socsys/fidentikit}.

\printbibliography

\appendix
\section*{Appendix}

\section{Dashboard View}
\vspace*{-1em}
\label{appendix:regex}
\begin{figure}
    \centering
\includegraphics[width=0.8\linewidth]{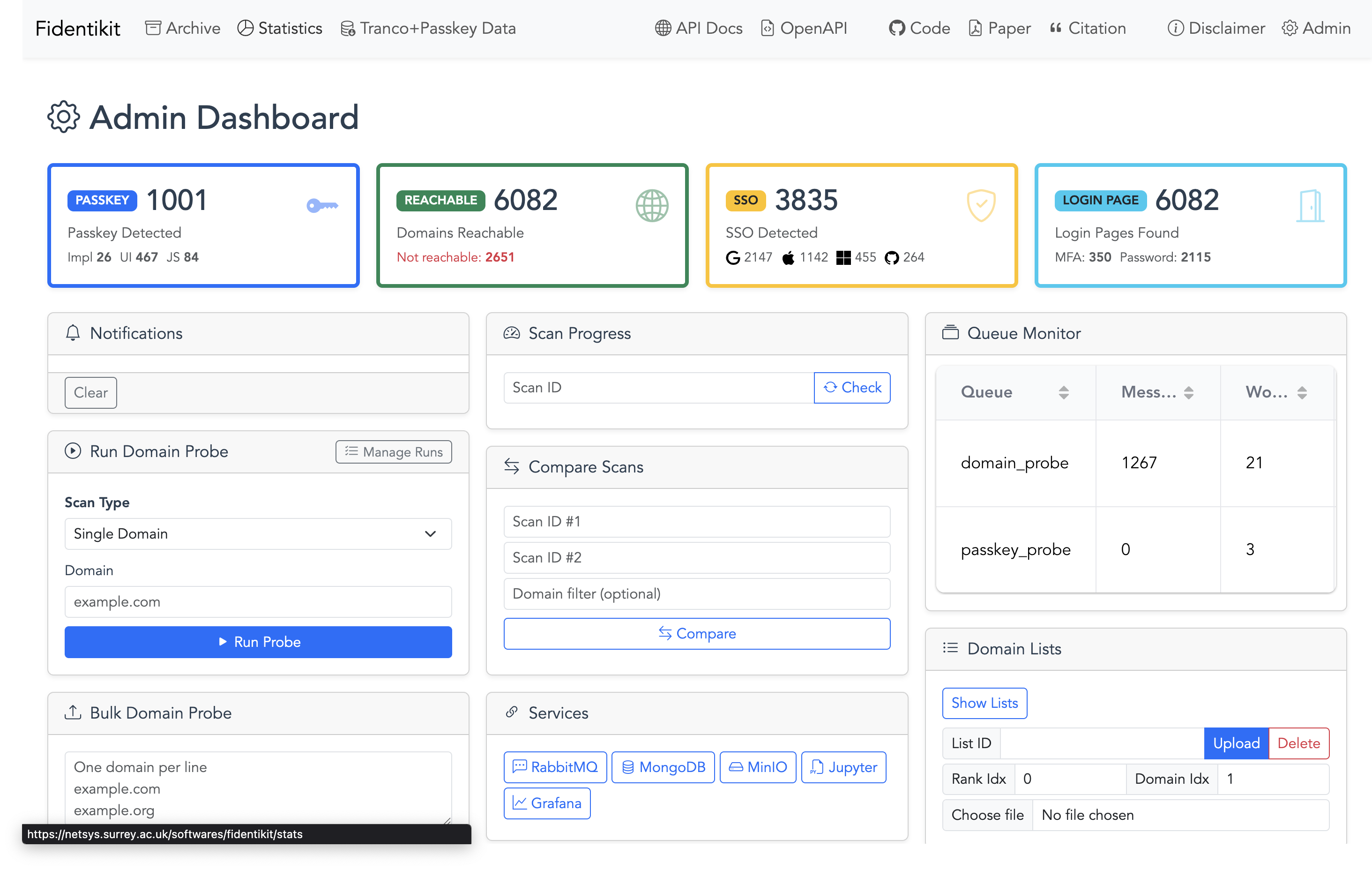}
\vspace{-0.2cm}
    \caption{Admin Dashboard}
    \label{fig:dashboard}
    \vspace*{-1.5em}
\end{figure}

\section{Detection Heuristics Catalogue}
\label{appendix:heuristics}

This appendix provides the complete specification of detection heuristics implemented in Fidentikit. We organise them by detection class.
\vspace*{-1em}
\subsection{JavaScript Pattern Detection}
\label{appendix:js-patterns}
Table~\ref{tab:js-patterns} lists the regular expressions used to detect WebAuthn implementation patterns in inline and external JavaScript. These patterns target actual passkey functionality rather than mere API availability.

\begin{table*}[htbp]
\centering
\caption{JavaScript WebAuthn Implementation Detection Patterns}
\label{tab:js-patterns}
\small
\begin{tabular}{llp{7.5cm}}
\toprule
\textbf{ID} & \textbf{Confidence} & \textbf{Pattern / Description} \\
\midrule
JS-1 & HIGH & \texttt{navigator\textbackslash.credentials\textbackslash.create\textbackslash s*\textbackslash(\textbackslash s*\textbackslash\{[\textbackslash s\textbackslash S]*?publicKey\textbackslash s*:} \\
& & WebAuthn credential creation with publicKey options \\
\midrule
JS-2 & HIGH & \texttt{navigator\textbackslash.credentials\textbackslash.get\textbackslash s*\textbackslash(\textbackslash s*\textbackslash\{[\textbackslash s\textbackslash S]*?publicKey\textbackslash s*:} \\
& & WebAuthn credential retrieval with publicKey options \\
\midrule
JS-3 & HIGH & \texttt{PublicKeyCredential\textbackslash.isUserVerifyingPlatformAuthenticatorAvailable} \\
& & Platform authenticator availability check \\
\midrule
JS-4 & HIGH & \texttt{PublicKeyCredential\textbackslash.isConditionalMediationAvailable} \\
& & Conditional mediation (autofill) support check \\
\midrule
JS-5 & MEDIUM & \texttt{\textbackslash.getCredential\textbackslash s*\textbackslash(\textbackslash s*\textbackslash\{[\textbackslash s\textbackslash S]*?type\textbackslash s*:\textbackslash s*['"]public-key['"]} \\
& & Credential API with public-key type \\
\midrule
JS-6 & MEDIUM & \texttt{authenticatorAttachment\textbackslash s*:\textbackslash s*['"]platform['"]} \\
& & Platform authenticator configuration \\
\midrule
JS-7 & MEDIUM & \texttt{authenticatorAttachment\textbackslash s*:\textbackslash s*['"]cross-platform['"]} \\
& & Roaming authenticator (security key) configuration \\
\midrule
JS-8 & MEDIUM & \texttt{userVerification\textbackslash s*:\textbackslash s*['"]preferred|required['"]} \\
& & User verification requirement configuration \\
\midrule
JS-9 & MEDIUM & \texttt{"challenge"\textbackslash s*:\textbackslash s*["'][A-Za-z0-9+/=]+["']} \\
& & WebAuthn challenge data (Base64-encoded) \\
\midrule
JS-10 & MEDIUM & \texttt{"publicKey"\textbackslash s*:\textbackslash s*\textbackslash\{[\textbackslash s\textbackslash S]*?"challenge"\textbackslash s*:} \\
& & PublicKey credential options object \\
\midrule
JS-11 & MEDIUM & \texttt{residentKey\textbackslash s*:\textbackslash s*['"]required|preferred['"]} \\
& & Discoverable credential (passkey) requirement \\
\bottomrule
\end{tabular}
\end{table*}

\subsection{UI Element Detection Patterns}
\label{appendix:ui-patterns}

Table~\ref{tab:ui-selectors} specifies CSS selectors and text patterns used for UI-based passkey detection. We filter social media icons to reduce false positives.

\begin{table*}[htbp]
\centering
\caption{UI Element Detection Selectors and Patterns}
\label{tab:ui-selectors}
\small
\begin{tabular}{llp{8cm}}
\toprule
\textbf{ID} & \textbf{Type} & \textbf{Selector / Pattern} \\
\midrule
UI-1 & Button Text & \texttt{/(passkey|sign.in.with.passkey|continue.with.passkey|use.passkey)/i} \\
UI-2 & Biometric Text & \texttt{/(fingerprint|face.?id|touch.?id|biometric|windows.?hello)/i} combined with \texttt{/(sign.?in|log.?in|login|continue)/i} \\
UI-3 & Data Attributes & \texttt{[data-webauthn], [data-passkey], [data-credential]} \\
UI-4 & Auth Method Attr & \texttt{[data-authentication-method="passkey"], [data-auth-type="passkey"]} \\
UI-5 & Credential Inputs & \texttt{input[autocomplete="webauthn"], input[type="publickey"]} \\
UI-6 & ARIA Labels & \texttt{button[aria-label*="passkey"], [role="button"][aria-label*="security key"]} \\
UI-7 & Images/Icons & \texttt{img[alt*="passkey"], svg[aria-label*="fingerprint"]} \\
\bottomrule
\end{tabular}
\end{table*}

\subsection{Text Keyword Patterns}
\label{appendix:keywords}

Table~\ref{tab:keywords} lists the keyword patterns used for text-based passkey detection, organised by confidence level.

\begin{table}[htbp]
\centering
\caption{Passkey Text Detection Keywords}
\label{tab:keywords}
\small
\begin{tabular}{llp{4.5cm}}
\toprule
\textbf{ID} & \textbf{Confidence} & \textbf{Pattern} \\
\midrule
KW-1 & HIGH & \texttt{sign\textbackslash s+in\textbackslash s+with\textbackslash s+passkey} \\
KW-2 & HIGH & \texttt{login\textbackslash s+with\textbackslash s+passkey} \\
KW-3 & HIGH & \texttt{use\textbackslash s+passkey} \\
KW-4 & HIGH & \texttt{continue\textbackslash s+with\textbackslash s+passkey} \\
KW-5 & HIGH & \texttt{passkey\textbackslash s+authentication} \\
KW-6 & MEDIUM & \texttt{passkey} (standalone) \\
KW-7 & MEDIUM & \texttt{webauthn} \\
KW-8 & LOW & \texttt{biometric\textbackslash s+login} \\
KW-9 & LOW & \texttt{passwordless\textbackslash s+login} \\
KW-10 & LOW & \texttt{login\textbackslash s+without\textbackslash s+password} \\
\bottomrule
\end{tabular}
\end{table}


\subsection{Third-Party Identity Provider Detection}
\label{appendix:idp-rules}

Table~\ref{tab:idp-rules} summarises the OAuth flow detection rules for major identity providers. Detection combines network request interception with SDK signature matching.

\begin{table*}[htbp]
\centering
\caption{Third-Party Identity Provider Detection Rules}
\label{tab:idp-rules}
\small
\begin{tabular}{llll}
\toprule
\textbf{Provider} & \textbf{Domain Pattern} & \textbf{Path Pattern} & \textbf{SDK Signatures} \\
\midrule
Google & \texttt{\^{}accounts\textbackslash.google\textbackslash.com\$} & \texttt{/gsi/select}, \texttt{/oauth2} & \texttt{gapi.auth2}, \texttt{google.accounts.id} \\
Apple & \texttt{\^{}appleid\textbackslash.apple\textbackslash.com\$} & \texttt{/auth/authorize} & \texttt{AppleID.auth} \\
Microsoft & \texttt{\^{}login\textbackslash.(live|microsoftonline)\textbackslash.com\$} & \texttt{/oauth} & \texttt{msal.js}, \texttt{@azure/msal-browser} \\
GitHub & \texttt{\^{}github\textbackslash.com\$} & \texttt{/login/oauth} & -- \\
Facebook & \texttt{facebook\textbackslash.com\$} & \texttt{/dialog/oauth} & \texttt{FB.login} \\
\bottomrule
\end{tabular}
\end{table*}

\end{document}